\documentclass[aps,twocolumn]{revtex4}
\usepackage{epsfig}
\newcommand{\be}{\begin{equation}}
\newcommand{\ee}{\end{equation}}
\newcommand{\bea}{\begin{eqnarray}}
\newcommand{\eea}{\end{eqnarray}}
\newcommand{\ba}{\begin{array}}
\newcommand{\ea}{\end{array}}

\begin{document}

\title{Violating conformal invariance: Two-dimensional clusters grafted to wedges, cones, and branch points of Riemann surfaces}

\author{Hsiao-Ping Hsu, Walter Nadler, and Peter Grassberger}
\affiliation{John-von-Neumann Institute for Computing, Forschungszentrum
J\"ulich, D-52425 J\"ulich, Germany}
                                                                                
\date{\today}
\begin{abstract}
Lattice animals are one of the few critical models in statistical mechanics violating conformal invariance.
We present here simulations of 2-d site animals on
square and triangular lattices in non-trivial geometries. The simulations
are done with the newly developed PERM algorithm which gives very precise 
estimates of the partition sum, yielding precise values for the entropic
exponent $\theta$ ($Z_N \sim \mu^N N^{-\theta}$). In particular, we 
studied animals grafted to the tips of wedges with a wide range of angles
$\alpha$, to the tips of cones (wedges with the sides glued together), and to 
branching points of Riemann surfaces. The latter can either have $k$ sheets 
and no boundary, generalizing in this way cones to angles $\alpha > 360$
degrees, or can have boundaries, generalizing wedges. We find
conformal invariance behavior,
$\theta \sim 1/\alpha$, only for small angles ($\alpha \ll 2\pi$),
while $\theta \approx const -\alpha/2\pi$ for $\alpha \gg 2\pi$.
These scalings hold both for wedges and cones.
A heuristic (non-conformal) argument for the behavior at large $\alpha$
is given, and comparison is made with critical percolation.
\end{abstract}

\maketitle

Lattice animals (or polyominoes, as they are sometimes called in mathematics
\cite{Golomb}) are just clusters of connected sites on a regular lattice.
Such clusters play an important role in many models of statistical physics, 
and are considered as the standard model for randomly branched polymers
\cite{Lubensky}.
The difference between the animal model and other cluster models such as
percolation is that the clusters appear with non-trivial weights in the latter,
while every cluster with the same number $N$ of sites has the same weight
in the animal ensemble (this gives, more precisely, {\it site animals}; in 
this letter we shall not consider other models in the animal universality 
class, such as bond animals or lattice trees \cite{PERM}).
Moreover,
lattice animals are one of the few critical models in statistical mechanics violating conformal invariance \cite{Cardy,Miller93}.
However, the deeper reasons for this violation
as well as the consequences are still poorly understood.

On the one hand, the animal problem is similar to other models of statistical 
physics in allowing a field theoretic formulation \cite{Lubensky} 
and in showing anomalous 
scaling laws in space dimensions $d<d_c=8$, where $d_c$ is called the upper 
critical dimension. In particular, the number of animals with precisely 
$N$ sites attached to a given fixed site (the ``partition sum"; notice that we 
count here shifted animals as different) scales for large $N$ as 
\be
   Z_N \sim \mu^N N^{1-\theta}                       \label{ZN}
\ee
where $\mu$ is the non-universal {\it growth constant} and $\theta$ is a critical
exponent which is independent of the lattice type, but depends on the global 
geometry of the lattice. Similarly, the gyration radius scales as $R_N \sim N^\nu$
where the {\it Flory exponent} $\nu$ is universal and also independent of 
the geometry of the lattice.

On the other hand, the statistics of lattice animals has a number of rather 
unusual features. In particular, as mentioned above,
it is not conformally invariant \cite{Miller93}.
Conformal invariance gives rather strong constraints in two dimensions. First
of all, it gives restrictions on critical exponents. The fact that the Flory
exponent $\nu$ is not exactly known for 2-d lattice animals (while it is known
exactly for unbranched polymers and most other 2-d models), is a 
consequence of the lack of conformal invariance.

Secondly, and more closely related to the present work, conformal invariance
gives strong constraints on the entropic exponent $\theta$ in non-trivial 
geometries. The most thoroughly studied of such geometries are wedges and cones.
A wedge is a part of the plane bounded by two straight lines which intersect
at an angle $\alpha$. A cone is basically a wedge where the two boundaries are 
glued together. A wedge can be mapped onto a half plane by a conformal map,
while a cone can be mapped onto a punctuated plane. This implies that for
conformally invariant theories the entropic exponent is linear in $1/\alpha$
\cite{Cardy,DeBell85,Miller93},
\be
   \theta = a+b/\alpha \;.                               \label{conform}
\ee
For models in which one can study single clusters (such as self avoiding 
walks, percolation, or lattice animals) this applies to clusters grafted
at the tip of the wedge or cone, respectively.
While this equation was checked for self avoiding walks (linear polymers)
\cite{Guttmann84,DeBell85}, it was indeed found 
{\it not} to hold for lattice animals \cite{Miller93,DeBell85}. But 
the numerical results given in the latter papers were not sufficient to 
suggest any alternative behaviour.

%Fig. 1
\begin{figure} [t]
  \begin{center}
    \psfig{file=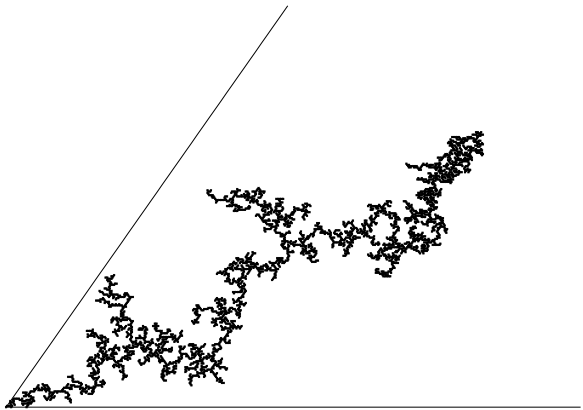,width=6cm,angle=270}
   \caption{Typical cluster for a wedge with $\alpha=\pi/3$ on a triangular lattice;
   $N=3500$.}
\label{fig-animal}
\end{center}
\end{figure}

In the present letter we apply the PERM (Pruned-Enriched Rosenbluth Method)
strategy which was recently adapted to lattice animals \cite{PERM}. It 
is a recursively implemented sequential Monte Carlo method with re-sampling
which starts off by growing percolation clusters, re-weighs them according 
to the animal ensemble, and applies cloning and pruning to achieve approximate
importance sampling. It is the most efficient algorithm for simulating animals
and lattice trees known today. In particular, it provides very precise 
estimates of the partition sum, which then allows to estimate $\theta$ by 
means of Eq.(\ref{ZN}). In the following, the maximal sizes of animals varied 
between $N_{\rm max} = 1000$ and $N_{\rm max} = 4000$.

The results presented below are based on simulations on the square and 
on the triangular lattices. While the growth constant for the square lattice 
was taken from \cite{PERM}, we had to perform new simulations on the full
lattice to obtain it for the triangular lattice. We obtained 
$\ln \mu_{\rm triang} = 1.6454139(18)$ (using the fact that $\theta=1$ on a 
full 2-d lattice).

Wedges on the square lattice are most easily obtained by placing one edge
along one of the coordinate axes (say the positive x-axis), and taking the 
angle $\alpha$ such that $\tan\alpha = n/m$ with $n$ and $m$ being integers.
Alternatively, one can place the x-axis along the center of the wedge, and 
use $\tan(\alpha/2) = n/m$ (``symmetrical wedge"). While these constructions
cannot be used e.g. for wedges of 30, 60, or 120 degrees, the latter can be 
obtained for triangular lattices,
see Fig.~\ref{fig-animal}. We checked that the standard and the
symmetrical wedge gave the same $\theta$ for 90 degrees (within error bars),
and that the square and triangular lattices gave the same results
for 180 degrees. Wedges with 360 degrees are obtained by excluding a single 
half line.

As we said, cones are obtained by gluing together the two edges of a
wedge. This can of course only be done if the lattices  agree along the 
two edges and if this does not introduce a line of defects, which strongly 
restricts the possible angles. For the square
lattice, only $\alpha=90, 180,$ and 270 degrees are possible, while 60, 120,
240, and 300 degrees are also possible for the triangular lattice. 
Enumeration data for cones with other angles are given in \cite{Miller93}, 
but it is not clear how they were obtained. In any case we checked that 
defect lines per se have an effect on $\theta$, by simulating 
animals grafted to points on a defect line.

To simulate angles larger than $2\pi$, we used multi-sheeted Riemann surfaces
with branch points at the origin where the animal is grafted. A branch
point where $k$ sheets meet is essentially a cone with angle $2k\pi$. Wedges
with $\alpha>2\pi$ are then obtained by cutting out the corresponding domain
from the surface.

 \begin{table}
 \begin{center}
 \caption{Entropic critical exponents for 2-d lattice animals grafted to the tip 
    of a wedge with angle $\alpha$.}
   \label{table2}
 \begin{ruledtabular}
 \begin{tabular}{rcl}
 $\alpha$ &        $\theta$      &  comment\\
\hline
 $\arctan(1/6)$ &  $16.74  \pm .05 $    &  square lattice\\
 $\arctan(1/5)$ &  $14.241 \pm .027$    &  square lattice\\
 $\arctan(1/4)$ &  $11.741 \pm .025$    &  square lattice\\
 $\arctan(1/3)$ &  $ 9.257 \pm .016$    &  square lattice\\
 $\arctan(1/2)$ &  $ 6.826 \pm .008$    &  square lattice\\
 $\pi/6     $ &  $ 6.204 \pm .008$    &  triangular lattice\\
 $\pi/4     $ &  $ 4.566 \pm .006$    &  square lattice\\
 $\pi/3     $ &  $ 3.739 \pm .006$    &  triangular lattice\\
 $\pi/2     $ &  $ 2.903 \pm .007$    &  square lattice\\
 $3\pi/4    $ &  $ 2.316 \pm .004$    &  square lattice\\
 $\pi$        &  $      2.0      $    &  exact \\
 $5\pi/4    $ &  $ 1.788 \pm .006$    &  square lattice\\
 $3\pi/2-\arctan(1/2) $ &  $ 1.718 \pm .006$    &  square lattice\\
 $3\pi/2    $ &  $ 1.622 \pm .007$    &  square lattice\\
 $7\pi/4    $ &  $ 1.478 \pm .007$    &  square lattice\\
 $2\pi      $ &  $ 1.354 \pm .008$    &  square lattice\\
 $3\pi      $ &  $ 0.790 \pm .02 $    &  square, 2-sheeted b. p.\\
 $4\pi     $  &  $ 0.358 \pm .02 $    &  square, 2-sheeted b. p.\\
 $6\pi      $ &  $-0.660 \pm .02 $    &  square, 3-sheeted b. p.\\
 $8\pi      $ &  $-1.678 \pm .03 $    &  square, 4-sheeted b. p.\\
 $10\pi$      &  $-2.670 \pm .05 $    &  square, 5-sheeted b. p.\\ \hline
 \end{tabular}
 \end{ruledtabular}
 \end{center}
\end{table}

 \begin{table}
 \begin{center}
 \caption{Entropic critical exponents for 2-d lattice animals grafted to the tip
    of a cone with angle $\alpha$.}
   \label{table1}
 \begin{ruledtabular}
 \begin{tabular}{rcl}
 $\alpha$ &        $\theta$      &  comment\\
\hline
 $\pi/3$  &  $2.5905 \pm .0025$  &  triangular lattice\\
 $\pi/2$  &  $2.1155 \pm .0025$  &  square lattice\\
 $2\pi/3$ &  $1.8570 \pm .0027$  &  triangular lattice\\
 $\pi$    &  $1.5490 \pm .0028$  &  square lattice\\
 $4\pi/3$ &  $1.3445 \pm .0025$  &  triangular lattice\\
 $3\pi/2$ &  $1.2549 \pm .0028$  &  square lattice\\
 $2\pi $  &  $        1.0     $  &  exact\\
 $4\pi $  &  $-0.011 \pm .007 $  &  square, 2-sheeted b. p.\\
 $6\pi $  &  $-1.012 \pm .009 $  &  square, 3-sheeted b. p.\\
 $8\pi $  &  $-2.020 \pm .013 $  &  square, 4-sheeted b. p.\\
 $10\pi$  &  $-3.031 \pm .023 $  &  square, 5-sheeted b. p.\\
 $12\pi$  &  $-4.04  \pm .04  $  &  square, 6-sheeted b. p.\\
 $14\pi$  &  $-5.045 \pm .05  $  &  square, 7-sheeted b. p.\\
 $16\pi$  &  $-6.06  \pm .06  $  &  square, 8-sheeted b. p.\\
 $20\pi$  &  $-8.045 \pm .07  $  &  square, 10-sheeted b. p.\\\hline
 \end{tabular}
 \end{ruledtabular}
 \end{center}
\end{table}

%Fig. 2
\begin{figure}
  \begin{center}
    \psfig{file=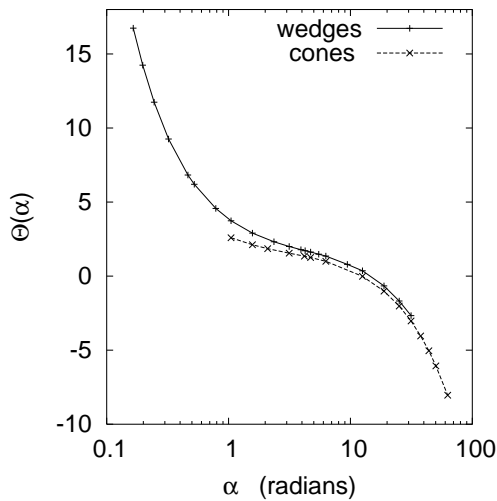,width=6.9cm,angle=270}
   \caption{Values of the entropic exponent $\theta$ for wedges (upper curve)
    and cones (lower curve) with angle $\alpha$, plotted against $\alpha$.
    Values for $\alpha > 2\pi$ were obtained using multi-sheeted Riemann
    surfaces. Statistical and systematic errors are much smaller than
    the sizes of the symbols.}
\label{fig-wedges}
\end{center}
\end{figure}

There are a number of implementation details which have to be specified. 
For instance, one has to specify for wedges whether a site on the boundary 
can be occupied or not. For a cone one might specify whether the two 
boundaries are identified, or whether they are one lattice unit apart.
Finally, for Riemann surfaces the site at the origin can have $\cal N$ 
neighbours ($\cal N$ is the coordination number, ${\cal N}=4$ for the square 
lattice and ${\cal N}=6$ for the triangular one), all of them on one sheet.
Or it can have $k{\cal N}$ neighbours, occupying all $k$ sheets. We checked
in each case several of these alternatives. As expected, they gave different 
results for finite $N$, but they led to the same scaling behaviours. 

Our final results are given in Tables 1 and 2, and are also shown in Fig.~\ref{fig-wedges}.
Notice that there are two values which are exact: $\theta=1$ for cones
with angle $2\pi$, and $\theta=2$ for wedges with $\alpha=\pi$ \cite{Lookman}.
For $\alpha \approx \pi$ our data agree very well with those of \cite{DeBell85},
but for smaller $\alpha$ the latter data seem to be systematically
too low: The value $\theta = 5.5\pm 0.1$ for $\alpha = 30$ degrees cited in 
\cite{DeBell85}, e.g., is
seven standard deviations below our value $6.204\pm 0.008$. A detailed 
comparison with the cone data of \cite{Miller93} is less straightforward. For 
those angles where cones without defect lines exist, the agreement is excellent.
For those where there should have been defect lines ($127^0,143^0, 233^0$), 
the data of \cite{Miller93} seem to be too high by two to four standard deviations 
(when compared to smooth interpolations of our data), as should be expected 
from the above discussion.

Both data sets (wedges and cones) confirm that Eq.(\ref{conform}) does not hold. 
But the wedge data (Table~1 and Fig.~\ref{fig-wedges}) show very clearly that Eq.(\ref{conform})
{\it does hold asymptotically} for $\alpha \to 0$:
\be
   \theta(\alpha) \approx 1.35 + b_{\rm wedge}/\alpha \qquad {\rm with}\quad b_{\rm wedge}=2.543\pm 0.020.
\ee
A similar scaling is also compatible with the cone data, although there the error 
on $b$ is much larger since we could not go to sufficiently small angles:
$b_{\rm cone} = 1.4\pm 0.1$.
On the other hand, both data sets indicate that $\theta$ increases linearly 
for large angles,
\be
   \theta(\alpha) \sim \alpha \qquad {\rm for} \quad \alpha \to\infty.
                   \label{linear}
\ee
Moreover, the coefficient of proportionality seems to be exactly the same in
both cases, 
\be
   \lim_{\alpha \to \infty} {\theta(\alpha)\over\alpha} = -{1\over 2\pi}\;(1.00\pm 0.03). 
\ee

%Fig. 3
\begin{figure}
  \begin{center}
    \psfig{file=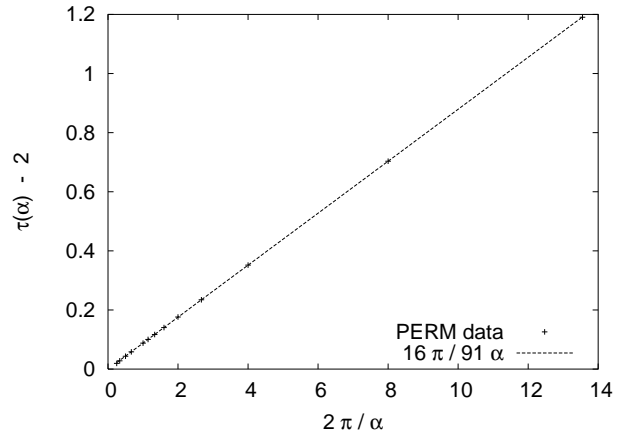,width=5.9cm,angle=270}
   \caption{Values of the exponent $\tau-2$ for critical percolation clusters, grafted
    on wedges with angle $\alpha$. The points (whose error bars are smaller than the 
    point size) are from simulations for site percolation on the square lattice, the 
    straight line is the theoretical prediction $\tau-2 = 16\pi / 91\alpha$.}
\label{fig-percol}
\end{center}
\end{figure}

In order to understand this behaviour heuristically, we now compare lattice 
animals (which are in the universality class of subcritical percolation) 
to critical percolation. If we use a standard cluster growth algorithm 
like the Leath algorithm \cite{Leath} or the depth-first algorithm of 
Swendsen and Wang \cite{Swendsen2} to grow critical or slightly subcritical
percolation clusters, the probability to reach a size $\geq N$ 
decreases like a power \cite{Stauffer92}, 
\be
   P_N \sim N^{2-\tau} f((p_c-p) N^\sigma).      \label{P}
\ee
Here, $p$ is the wetting probability, $p_c$ its critical value, $\tau$ and 
$\sigma$ are critical exponents, and $f(x)$ decreases exponentially for 
$x\to \infty$. The critical exponents can be related to other, more standard,
exponents, e.g. \cite{Stauffer92}
\be
   \tau=3-\gamma\sigma .
\ee
The ansatz Eq.(\ref{P}) holds both for clusters grown in the bulk, and for 
clusters grown near a surface. In the latter case, $p_c$ and the exponent 
$\sigma = 36/91$
are the same as in the bulk (and independent of the shape of the surface), while 
$\tau$ and $f(x)$ do depend on the surface. In particular it is known \cite{Cardy}
that $\gamma = 25/12$ for 2-d clusters attached to a plane wall, giving 
$\tau(\alpha = \pi) -2 = 16/91$.

From conformal invariance we expect that $\tau(\alpha)$ is a linear function of $1/\alpha$. 
On the other hand, we expect that $\tau \to 2$ for $\alpha \to \infty$. The reason 
is very simple: For $\alpha \to\infty$, the chances that the growth will stop
at any finite $N$ will go to zero, since there are ever more possible directions 
for growth. Thus we expect 
\be
   \tau(\alpha) = 2 + {16\pi\over 91\alpha}.
\ee
From Fig.~\ref{fig-percol} we see that this is in excellent agreement with simulations of clusters
starting at the tip of a wedge. Notice that the angles shown in Fig.~\ref{fig-percol} extend up
to $8\pi$, verifying thereby that branch points of Riemann surfaces can be treated
like tips of wedges.

For lattice animals one could also try to use Eq.(\ref{P}), this time with 
$p\ll p_c$, but this would not lead to any useful prediction. Thus one has to 
proceed differently. Our main assumption is that clusters will grow 
essentially into an angular region of size $\Delta\alpha = O(1)$. Much larger
angular ranges will also occur, but only with very low probability. For $\alpha
\gg \Delta\alpha$, i.e. $k\equiv \alpha / \Delta\alpha \gg 1$, one has thus 
essentially $k$ independent clusters. If one assumes that all these subclusters 
have roughly the same size, one expects
\be
   Z_N(k\alpha) \approx [Z_{N/k}(\alpha)]^k .
\ee
Together with Eq. (\ref{ZN}) this gives $\theta \propto \alpha$ for large $\alpha$. 
This argument explains also why the proportionality constant is the same for 
wedges and cones, but it does not explain its numerical value of $1/2\pi$, i.e. 
it does not explain why $\theta$ decreases by exactly one unit when $\alpha$ 
increases by $2\pi$.

If lattice animals are not conformally invariant but in some way ``covariant", 
one might expect a simple analytical formula for $\theta(\alpha)$. We therefore 
tried to find such fits. Simple ansatzes like $\theta = a+b\alpha+c/\alpha$ were
not successful. The simplest acceptable fits were obtained with Pad\'e approximates 
of the form $\theta = (a+b\alpha+c\alpha^2+d\alpha^3)/(\alpha+e\alpha^2)$. But the 
coefficients $a$ to $e$ looked rather uninspiring, and such an ansatz seems already 
too complicated for being ``natural".

\bigskip

In summary, we have presented very high statistics simulations of clusters grafted 
to the tips of cones and wedges. For critical percolation clusters these simulations 
were in full agreement with predictions from conformal invariance. But for lattice 
animals (subcritical percolation clusters) they agreed with the conformal invariance behavior only
in the limit of small angles. For large angles another simple behaviour was found
and explained by assuming distant angular regions to be essentially independent.
Our results clearly show consequences of the violation of conformal invariance in the lattice animals model. Nevertheless, the results also suggest that for small angles conformal invariance may still hold in some approximate way or that some generalized invariance might exist.

\end{document}